\documentclass[useAMS]{mn2e}
\usepackage{graphicx}
\usepackage{epsfig}

\def\ltsima{$\; \buildrel < \over \sim \;$}
\def\lsim{\lower.5ex\hbox{\ltsima}}
\def\gtsima{$\; \buildrel > \over \sim \;$}
\def\gsim{\lower.5ex\hbox{\gtsima}}

\newcommand{\be}{\begin{equation}}
\newcommand{\en}{\end{equation}}

\def\msole {~M_{\odot}}

\begin{document}
\title[XMM-Newton and Swift observations of the Type IIb SN 2011dh]{XMM-Newton and Swift observations of the Type IIb SN 2011dh in Messier 51}

\author[S. Campana \& S. Immler]{Sergio Campana$^{1,}$\thanks{E-mail: sergio.campana@brera.inaf.it}, 
Stefan Immler$^{2,3}$ \\
$^1$ INAF-Osservatorio Astronomico di Brera, Via Bianchi 46, I--23807, Merate (LC), Italy\\
$^2$ NASA/Goddard Space Flight Center, Astrophysics Science Division, Code 662, Greenbelt, MD 20771, USA\\ 
$^3$ Department of Astronomy, University of Maryland, College Park, MD 20742, USA
}

\maketitle

\begin{abstract}
The Type IIb SN 2011dh exploded in the nearby galaxy M51 (the Whirlpool Galaxy) and provides us 
with one of the best laboratory to study early high energy emission from SNe. 
We give here a comprehensive view of the X--ray properties of SN 2011dh from the analyses of two pointed XMM-Newton 
early observations as well as of the full Swift X--ray Telescope (XRT) dataset (163 ks). Due to the 
high XMM-Newton throughput, we were able to satisfactorily fit the X--ray spectrum with two hot diffuse gas 
components including an additional absorption component to our Galaxy. A power law model provided a 
worse description of the data. In addition, the early Swift XRT light curve hints of a flux excess 
at early times ($\lsim 3$ d), consistent with the adiabatic cooling of stellar's photosphere a few days after the shock breakout.
\end{abstract}


\section{Introduction}

Massive stars ($M\gsim 8\msole$) end their lives with a (core-collapse) supernova (SN) explosion. 
It is notoriously difficult to detect X--ray emission from SNe. Dominant X--ray emission mechanisms comprise 
an energetic shock break out soon after the explosion (Campana et al. 2006; Soderberg et al. 2008) or the interaction 
of the shock front arising from the SN explosion with the circumstellar medium (CSM).
In particular, collisions of the ejected material with the CSM (made by material ejected from the 
progenitor prior to explosion) can heat up matter to $T\sim 10^9$ K, leading to X--ray emission. 
A reverse shock will also form propagating inward with a temperature $T\sim 10^7$ K. 
Alternatively, X--ray emission can arise from Compton down-scattered gamma--rays of radioactive decay in the ejecta, 
or from inverse Compton scattering of relativistic electrons on photospheric UV photons produced during the outburst.

Nearby SNe represent an ideal target for X--ray studies. On May 31.893, 2011 UT a SN was discovered by several groups
in the Whirlpool galaxy M51 ($d\sim 8.4$ Mpc, Feldmeier, Ciardullo \& Jacoby 1997). 
The SN was classified as a Type IIb (Arcavi et al. 2011;
Marion et al. 2011) and named SN 2011dh (Griga et al. 2011; Silverman, Filippenko \& Cenko 2011).
Given its closeness attempts to identify a progenitor in pre-explosion Hubble Space Telescope (HST) images have 
been carried out. A possible progenitor has been identified even if it is not certain (Maund et al. 2011; van Dyk et al. 2011).
Photometry of this object is consistent with a yellow super-giant star. Along these lines Bersten et al. (2012) modeled the 
optical light curve and the photospheric velocity evolution based on a set of hydrodynamical models, finding a very large
progenitor radius of $R_*\sim 10^{13}$ cm. A compact progenitor was instead suggested by other authors (Soderberg et al. 2012; 
Arcavi et al. 2011). In this case the identification of the progenitor was questioned and the radius estimate was based on radio 
properties leading to a radius estimate of $R_*\sim 10^{11}$ cm.

SNe IIb may be divided into two sub-classes based on the radius of the progenitor and on its mass loss history 
(Chevalier \& Soderberg 2010). Compact progenitors give rise to fast shock waves ($\sim 0.1\,c$) and modulated radio light curves 
(SN cIIb); extended progenitors ($R_*\sim 10^{13}$ cm) are characterized by slower shock waves and smoother radio curves (SN eIIb).
Hydrodynamical light curve modeling gives strong evidence that SN 2011dh belongs to the class of extended progenitors 
($R\sim 200\,R_\odot$; Bersten et al. 2012).

Here we report on X--ray observations on SN 2011dh. We will first report on two XMM-Newton observations 7 and 11 d after the SN 
explosion (Section 2). In light of these observations we will reanalyse Swift X--ray Telescope (XRT) observations (Section 3), expanding 
the results reported in Soderberg et al. (2012). Section 4 is dedicated to discussion and conclusions.

\begin{table*}
\caption{XMM-Newton observations log}
\begin{center}
\begin{tabular}{cccc|cccc}
\hline
Obs.1& Exp. time&Filt. exp. time& Collected     & Obs.2  & Exp. time&Filt. exp. time & Collected\\
Jun 7.2 2011 &  (ks)  &  (ks)              &photons$^*$&Jun 11.2 2011&  (ks)           &  (ks)                  & photons$^*$\\  
\hline
pn             & 9.8    & 9.8                 & 719 ($75\%$)& pn        & 9.7             & 2.6 & 123 (67\%)\\
MOS1      & 10.6    & 10.6             & 165 ($79\%$)& MOS1 & 6.4             & 5.5 & 80 (75\%)\\
MOS2      & 10.7    & 10.7             & 181 ($67\%$)& MOS2 & 6.3             & 5.5 & 60 (76\%)\\
\hline
\end{tabular}
\end{center}
$^*$ Photons within the extraction region in the 0.2--12 keV energy range for the pn and 0.3--10 keV for the MOS.
In parenthesis we also reported the contribution of the source with respect to the estimated background.
\end{table*}

\begin{figure*}
\begin{center}
\includegraphics[width=16cm]{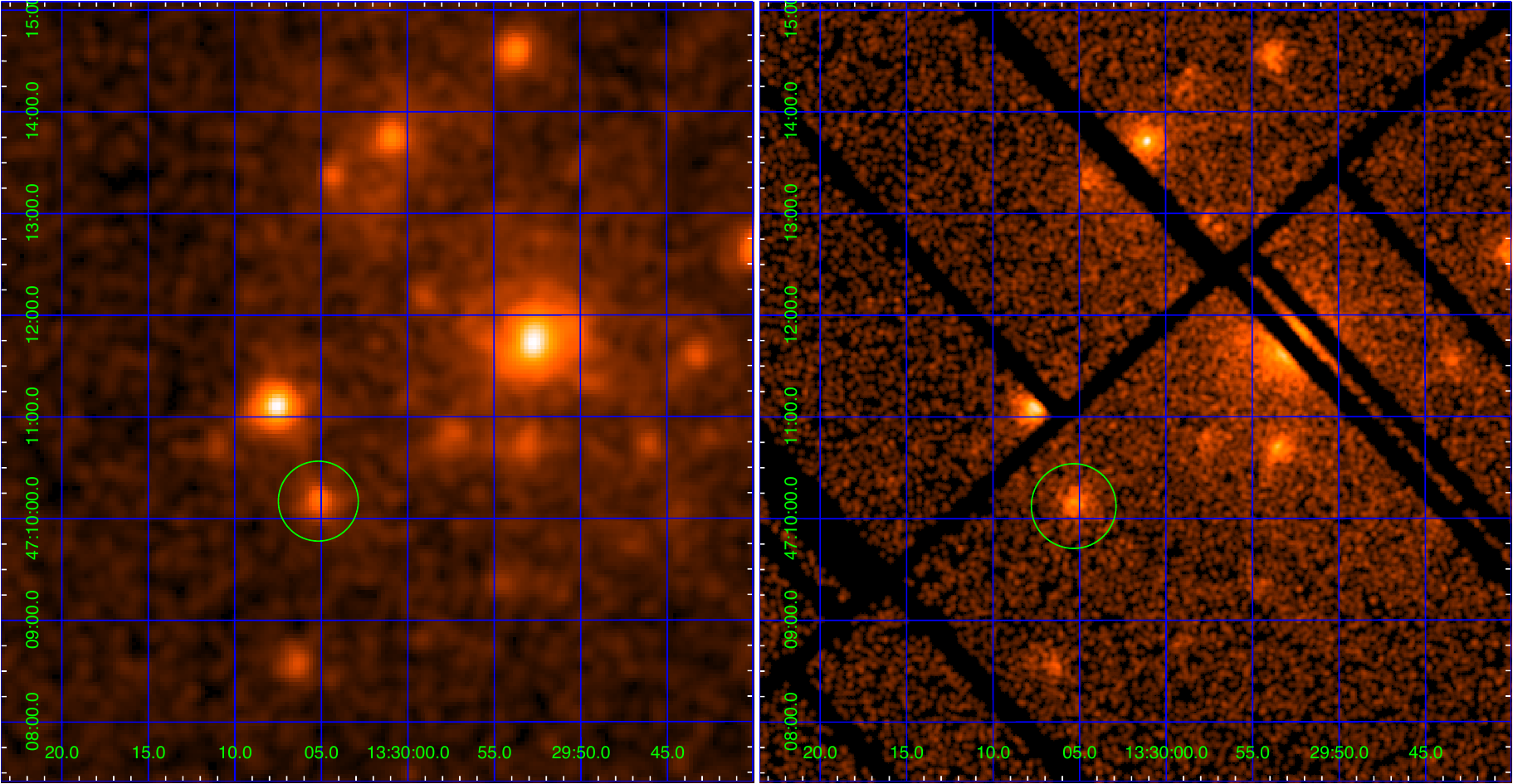}
\end{center}
\caption{Left: smoothed image of all the Swift XRT observations (163 ks). Right ascension on the $x$-axis and declination on the $y$-axis.
Right: smoothed image of the first XMM-Newton pn observation ($\sim 10$ ks). 
Images were smoothed with a Gaussian function with 3 pixel kernel radius.
SN2011dh is indicated in both images with a circle, that is also the extraction region adopted.}
\label{ima}
\vskip -0.1truecm
\end{figure*}

\section{XMM-Newton observations}

XMM-Newton observed M51 twice based on a DDT program (P.I. Campana). 
The first observation took place on June 7.207, 2011 UT and lasted $\sim 10$ ks (see Table 1). The thin filter
was used on all the X--ray instruments for all the observations. During the entire observing time the background remained low and 
we retained all the observing time. The second observation took place on June 11.196, 2011 UT for $\sim 10$ ks (see Table 1).
In this case a strong background flare affected the observation. We filtered the MOS observations requiring less than 10 c s$^{-1}$ on the 
entire instrument. For the pn we used a 30 c s$^{-1}$ threshold. The resulting exposure times are reported in Table 1.

Data from the first observation were extracted from a 500 pixel (25 arcsec) radius centered on source. The background was evaluated in 
two nearby regions both with 800 pixel radius on the same CCD surrounding the SN 2011dh location and free of sources and bad columns
(see Fig. \ref{ima}). Chandra images show that within this extraction region there is just a faint source to the south-west (Soderberg et al. 
2012). Given the brightness of SN 2011dh, contamination of the data is minimal and does not affect our analysis.
We collected 1065 and 263 photons for the first and second observation, respectively. Data were group to have at least 20 counts per energy bin
and {\tt rmf} and {\tt arf} files were generated using the latest calibration products. 

\begin{figure}
\begin{center}
\includegraphics[width=6cm,angle=-90]{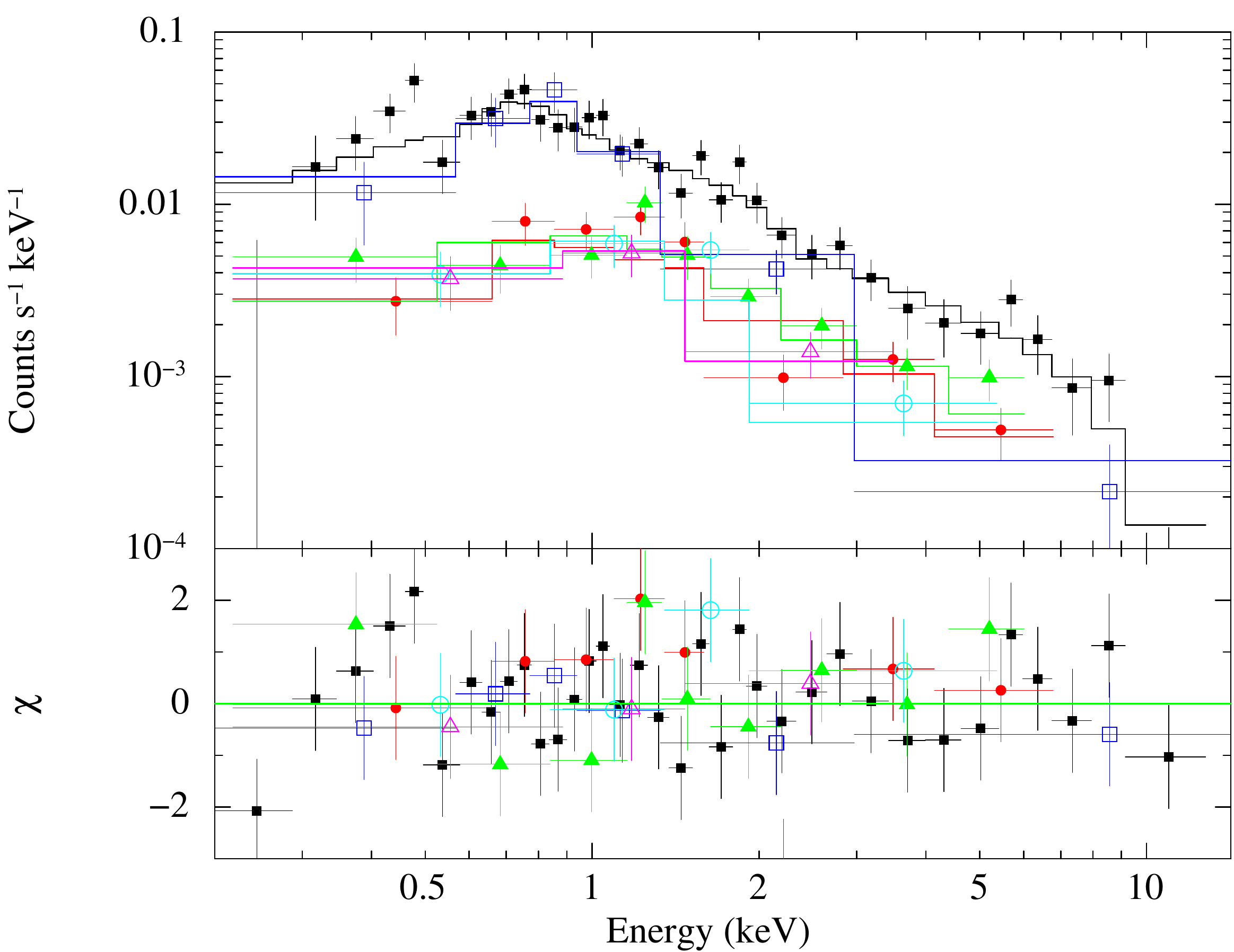}
\end{center}
\caption{XMM-Newton spectra of SN2011dh. Squares refer to pn data, circles to MOS1 data, and triangles to MOS2 data.
Filled markers indicate data from the first observation, open markers from the second observation (black: 1$^{\rm st}$ pn, red: 1$^{\rm st}$ MOS1,
green: 1$^{\rm st}$ MOS2; blue: 2$^{\rm nd}$ pn, light blue: 2$^{\rm nd}$ MOS1, magenta: 2$^{\rm nd}$ MOS2).
}
\label{spe}
\vskip -0.1truecm
\end{figure}

We fit together the six spectra, keeping the same absorption pattern for all of them 
(two {\tt TBABS} models, one fixed to the Galactic value of $1.8\times 10^{20}$ cm$^{-2}$; Kalberla et al. 2005)
and allowing for the same normalization 
constant for MOS1 and MOS2 in the two observations, while keeping the pn normalization fixed to one.
The second spectrum is softer than the first one. We tried a few single component models. The null hypothesis probabilities
are never larger than a few percent, indicating poor fits. We also tried a two more complex models.
A model consisting of a power law and an emission from hot diffuse gas ({\tt MEKAL}) provides an improvement 
($3\%$ according to the F-test) but the resulting intrinsic column density is extremely large and not consistent 
with optical data (Arcavi et al. 2011; Ritchey \& Wallerstein 2012, see also below).
A model with two hot thermal gas components gives a better fit to the data
(null hypothesis probability of $15\%$) and an improvement over a single {\tt MEKAL} of $2.7\,\sigma$ 
($0.7\%$, according to the F-test, Fig. \ref{spe}). 

Given the two best fit models we tried to fit the data without the presence of an intrinsic absorption component 
and evaluate the improve in the fit due to its inclusion with an F-test. In both cases the probability of a chance 
improvement  is $2\%$ ($2.3\,\sigma$). We also note that we assumed for the Galactic column density contribution 
a value of $1.8\times 10^{20}$ cm$^{-2}$ (Kalberla et al. 2005) which is larger than the values derived from the 
Dickey \& Lockman (1990) survey of $1.6\times 10^{20}$ cm$^{-2}$ or from the Stark et al. (1992) survey of 
$1.5\times 10^{20}$ cm$^{-2}$. This indicate that the estimated significance of our intrinsic component is a tight lower limit.

In all models a non-negligible absorption component in addition to the Galactic one is needed. This indicates  
the existence a small but non-zero intrinsic absorption, possibly arising from the absorbing shell between the 
forward and reverse shock. The number of collected photons is too low to assess if a variation of the 
column density is occurring among the two observations.

\begin{table*}[htb]
\caption{XMM-Newton spectral fits.}
\begin{center}
\begin{tabular}{ccccc}
\hline
Model         & Column density    & 1$^{\rm st}$ Obs.              & 2$^{\rm nd}$ Obs.           &$\chi^2_{\rm red}$/dof \\ 
                     &$(10^{20}$ cm$^2)$& $\Gamma$/$k\,T$ (keV)& $\Gamma$/$k\,T$ (keV)& (n.h.p.)                           \\
\hline
Power-law &$6.4^{+3.9}_{-3.3}$& $1.6\pm0.2$                     & $2.0\pm0.3$                    & 1.4/58 (0.02)\\
Black body&$<0.9$                      &$0.39^{+0.06}_{-0.04}$   &$0.34^{+0.04}_{-0.02}$  & 3.4/58( $10^{-16}$)\\
Bremss.     &$2.7^{+2.3}_{-2.1}$&$14.9^{+26.1}_{-7.2}$    &$2.8^{+3.0}_{-1.2}$          & 1.5/58( 0.01)\\
Mekal         &$2.3^{+2.1}_{-1.9}$&$11.9^{+25.5}_{-4.7}$    &$3.0^{+2.4}_{-0.9}$            & 1.4/58( 0.02)\\
\hline
Mekal+pow&$67.2^{+27.5}_{-7.6}$&$0.22\pm0.01$            &$0.29^{+0.45}_{-0.15}$& 1.3/54 (0.07)\\
                     &                                     &$1.0\pm0.2$                   &$1.3\pm0.4$                    & \\
\hline
Mekal+Mekal&$4.7^{+5.5}_{-3.3}$&$>10.7$                          & $>2.4$                            & 1.2/54 (0.15)\\
                      &                                     &$0.35^{+0.22}_{-0.08}$&$0.59^{+0.26}_{-0.28}$& \\
\hline
\end{tabular}
\end{center}

Errors have been calculated for a $\Delta \chi^2=2.71$, corresponding to a $90\%$ confidence level for one parameter
of interest.

The column density reported in the Table is in addition to the Galactic column density ($N_H=1.8\times 10^{20}$ cm$^{-2}$)
and it likely is within the host galaxy.

Leaving free the metal abundance of the {\tt MEKAL} model, does not improve the fit, obtaining only an upper limit on 
the metallicity of $Z<3.9\,Z_\odot$.

\end{table*}

We adopted this model as our reference model (last row in Table 2). In this case the unabsorbed 0.5-10 keV flux is $2.4\times 10^{-13}$ erg cm$^{-2}$
s$^{-1}$ and $1.1\times 10^{-13}$ erg cm$^{-2}$ s$^{-1}$, respectively. These correspond to a 0.5-10 keV luminosities 
of $2.0\times 10^{39}$ erg s$^{-1}$  and $9.3\times 10^{38}$ erg s$^{-1}$, respectively (adopting a power law, fluxes and 
luminosities are $\sim 10\%$ lower).

\section {Swift X--ray Telescope observations}

\begin{figure}
\begin{center}
\includegraphics[width=6cm,angle=-90]{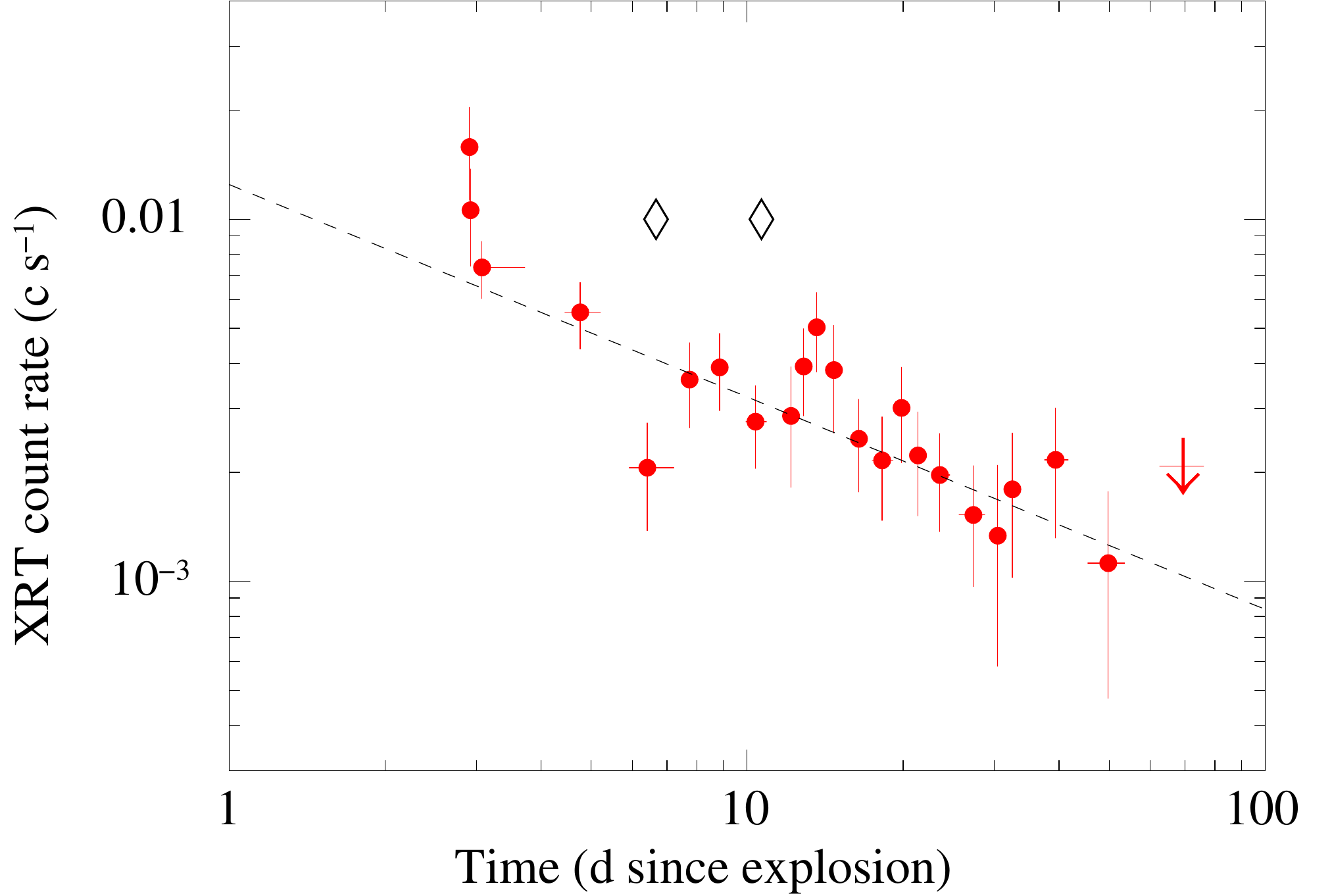}
\end{center}
\caption{Swift XRT light curve starting from the best estimate of the SN explosion time. 
The light curve has been obtained using the Leicester University user\_object tool, 
by adaptively summing up data to reach a $6\,\sigma$ detection. The two diamonds indicate the times of the two 
XMM-Newton observations.}
\label{lc}
\vskip -0.1truecm
\end{figure}

Swift promptly detected SN 2011dh about 3 days after the explosion (Margutti \& Soderberg 2011). Swift observed M51 112 times.
40 observations were  carried out before the SN 2011dh explosion. Soderberg et al. (2012) provided a $3\,\sigma$ upper limit at the 
SN position of $5.6\times 10^{-4}$ c s$^{-1}$ (converting to a 0.5--10 keV luminosity of $\sim 2\times 10^{38}$ erg s$^{-1}$).
The remaining 72 observations were targeted on SN 2011dh spanning the time period June 3 - August 15, 2011 for a total 
observing time of 163 ks (Fig. \ref{ima}).
The light curve was built using the UKSSDC light curve generator\footnote{http://www.swift.ac.uk/user\_objects/} paying particular care 
not to include nearby sources (Fig. \ref{lc}). Given the variable XRT background caused by CCD temperature variations, we prefer
not to rely on pre-explosions observations. 
Each point in the light curve has a significance of $6\,\sigma$. We assumed an explosion data of May 31.6, 
2011 UT, based on the analysis of Soderberg et al. (2012). This sets the origin of the time axis in the Fig. 3. 
The overall curve can be fit with a power law decay with index $-0.59\pm0.16$ (including the small uncertainty involved by the 
determination of the SN explosion of 0.2 d, see Fig. 1). The fit is acceptable with a reduced $\chi^2_{\rm red}=1.2$ (20 d.o.f.). 
One can notice a flux excess at the beginning of the observing time, possibly related a SN shock break out (Campana et al. 2006; 
Soderberg et al. 2008). At the beginning of the observing time, the first point in the X--ray light curve lies at $\sim 2.5\,\sigma$ from the model.
The X--ray behavior might also be consistent with the optical/UV light curve
(Arcavi et al. 2011).
A bump in the X--ray light curve, even if not statistically significant, can also be observed at $\sim 13$ d after 
the SN explosion. This is barely consistent with the peak of the Swift/UVOT light curve in the $U$ band, when the SN peak is clearly 
visible (Arcavi et al. 2011). 

We extracted two set of spectra. The first comprise the entire dataset. Assuming the same absorption pattern derived from XMM-Newton 
observations we can fit the 414 counts in the 0.3--10 keV energy band (wit only $71\%$ due to the source) with a power law model.
The best fit photon index is $\Gamma=1.48\pm0.18$. The reduced $\chi^2_{\rm red}=1.5$ with 18 d.o.f. ($7\%$ null hypothesis probability).
The average flux 0.3--10 keV unabsorbed flux is $1.2\times 10^{-13}$ erg cm$^{-2}$ s$^{-1}$ (giving a luminosity of $1.0\times10^{39}$ erg s$^{-1}$).
A double {\tt MEKAL} model improves only marginally the fit. This is expected because spectral evolution was observed, at least, across the
two XMM-Newton observations. 
A spectrum was also extracted taking data from the first five days, summing 30 ks of data and collecting 134 photons ($5\%$ background).
In this case the spectrum is harder, with a power law photon index of $\Gamma=1.1\pm0.3$ (reduced $\chi^2_{\rm red}=1.2$ for 11 d.o.f., $27\%$ n.h.p.)
The 0.3--10 keV unabsorbed flux is three times more than the average value. 

\section{Discussion}

X--ray absorption holds the potential of tracing the presence of metals through their impact on the X--ray spectrum.
XMM-Newton observations clearly indicate a non-negligible absorbing column density 
in addition to the Galactic contribution. This likely arise within M51. Assuming a solar distribution of elements and a solar metallicity 
the column density is $\log N_H=20.7^{+0.3}_{-0.5}$ ($90\%$ confidence level). 
This value can be compared to optical estimates. Arcavi et al. (2011) estimated $E(B-V)\lsim0.05$. This can be converted to 
$\log N_H\sim 20.5$, following Predehl \& Schmitt (1995). 
Data from the The H I Nearby Galaxy Survey (THINGS; Walter et al. 2008) yields at the position of SN 2011dh $\log N(H\, I)=20.6\pm0.1$. 
A more thorough analysis was carried out by Ritchey \& Wallerstein (2012). From echelle observation of the SN,  they derived Na I column 
densities of the multiple components and estimated  $\log N(H\,I)\sim 19.8-20.2$. 
The value derived from X--ray observations is slightly larger than this but still consistent. This would indicate that the surrounding medium 
has a metallicity similar to the solar value.

One question is if this instrinsic absorbing column density is intrinsic to the source or due to some intervening matter in M51 along the 
line of sight but far away from the SN site. M51 is a nearby galaxy and it is well studied in the X--ray band.  M51 hosts a number of ultra-luminous 
X--ray sources (ULXs) well studied and characterized (Dewangan et al. 2005; Yoshida et al. 2010). 
The two closest ULX sources to SN 2011dh are source 82 ($\sim 1$ arcmin to the north) and source 63 ($\sim 1.5$ arcmin to the north-west).
Both sources are variable. The column density to source 82 is at a level of  $\log N_H\sim 21$ (Yoshida et al. 2010).
We confirm this value with the analysis of our XMM-Newton data (assuming a power law spectrum).
Source 63 is fainter and the intrinsic column density (assuming a power law plus {\tt MEKAL} model) is consistent with zero (Dewangan et al. 2005).
In our data source 63 is a factor of $\sim 3$ fainter and we cannot confirm these findings.
We also note that the Type Ic SN 1994I occurred 2.3 arcmin away from SN 2011dh. No direct evidence for the presence of circumstellar matter
was found in X--ray data (namely too few photons to assess an increase over the Galactic column density value), but the X--ray emission was 
interpreted as due to the interaction of the blast wave with its surrounding circumstellar material (Immler, Wilson \& Terashima 2002).
 From this comparison we can assess that we have, at least, found strong indications that the intrinsic column density observed in SN 2011dh 
is due to circumstellar material, as expected for a supergiant progenitor (Bersten et al. 2012).

X--ray emission from SNe IIb has been observed in a very limited number of objects: SN 1993J (Chandra et al. 2009), SN 2001gd (P\'erez-Torres et al. 2005), 
SN 2001ig (Schlegel \& Ryder 2002), SN 2003bg (Soderberg et al. 2006), SN 2008ax (Roming et al. 2009),  and SN 2011dh.
SN 1993J is probably the best studied object. Using Rosat data, Zimmermann \& Aschenbach (2003) found that the intrinsic column densities is much 
higher than the Galactic column density. A deeper look came from ASCA data where Uno et al. (2002), fitting a two-component thermal-plasma model,
found that the low-temperature component has much higher column density than the high-temperature component. These high column densities 
in excess to the Galactic absorption were attributed to the absorption by an additional cool shell. For all the other X--ray detected SN IIb we have 
too few counts to assess if an intrinsic column density excess is present.
Thanks to the larger XMM-Newton throughput, we were able to assess that the X--ray spectrum consists of one soft and one hard 
emission components (modeled with the {\tt MEKAL} model), likely related to a reverse shock as in SN 1993J (Chandra et al. 2009; Nymark et al. 2006).
Our best fit two component thermal plasma model is then fully consistent with shock model physics from SN explosions, 
i.e., a hard thermal plasma component (at a few keV) that arises from the fast forward shock into the circumstellar material, 
and a soft thermal plasma component (below 1 keV) from the emergence of a reverse shock at lower speeds into the ejecta. 
Since the reverse shock is located behind the forward shock, it is expected to have a higher absorbing column density than the forward shock.
Our XMM-Newton observations clearly show a non-negligible absorbing column density in addition to the Galactic contribution.

Finally, a reanalysis of the Swift XRT light curve might indicate a flux excess at the beginning of the X--ray observations. 
Despite the fact that the statistical significance of this increase is low and the physical interpretation is challenging, 
this excess is likely due to the adiabatic cooling of stellar's photosphere a few days after the shock breakout.
The time of the X--ray excess is identical to the time of the optical excess described by Arcavi et al. (2011) and interpreted 
by them as the tail of the shock breakout emission. This might hint for a non-compact progenitor, i.e. yellow supergiant progenitor.

\section{Acknowledgments}
SC thanks Norbert Schartel for granting DDT XMM-Newton observations and Neil Gehrels and the Swift science operations team 
for approving and scheduling the Swift observational campaign. We thank the referee for comments that helped improving the manuscript.
This work made use of data supplied by the UK Swift Science Data Centre at the University of Leicester.
This work has been partially supported by the ASI grant I/004/11/0 and by the PRIN-MIUR grant 2009ERC3HT.

{}

\end{document}